\documentclass[traditabstract,letter]{aa} 
\usepackage{graphicx}
\usepackage{txfonts}
\usepackage{natbib}

\def\ltsima{$\; \buildrel < \over \sim \;$}
\def\lsim{\lower.5ex\hbox{\ltsima}}

\def\mpch{\mbox{$h^{-1}$Mpc}}

\def\deg{\ifmmode{^\circ}\else{$^\circ$}\fi}
\def\hGpc{\ifmmode{h^{-1}{\rm Gpc}}\else{$h^{-1}{\rm Gpc}$}\fi}
\def\hkpc{\ifmmode{h^{-1}{\rm kpc}}\else{$h^{-1}{\rm kpc}$}\fi}
\def\hMpc{\ifmmode{h^{-1}{\rm Mpc}}\else{$h^{-1}{\rm Mpc}$}\fi}
\def\hMsun{\ifmmode{h^{-1}M_\odot}\else{$h^{-1}M_\odot$}\fi}

\def\muK{\ifmmode{\mu{\rm{K}}}\else{$\mu$K}\fi}
\def\mum{\ifmmode{\mu{\rm{m}}}\else{$\mu$m}\fi}

\def\mrlogh{M_r-5\log h}

\begin{document}
   \title{Quantifying the coherent outflows of galaxies around voids in the SDSS DR7}

   \author{S.~G. Patiri
          \inst{1} \and J. Betancort-Rijo\inst{2}
          \fnmsep\inst{3} 
           \and F. Prada\inst{4}\fnmsep\inst{5}\fnmsep\inst{6}
          }

   \institute{IANIGLA-CONICET, Parque Gral San Mart\'in, C.C. 330, 
5500 Mendoza, Argentina\\ \email{spatiri@mendoza-conicet.gob.ar} 
\and Instituto de Astrof\'isica de Canarias, c/ Via L\'actea s/n, 
E-38200 Tenerife, Spain 
\and Departamento de Astrof\'isica, Universidad de La Laguna, 
E-38205 Tenerife, Spain 
\and Campus of International Excellence UAM+CSIC, Cantoblanco, 
E-28049 Madrid, Spain 
\and Instituto de F\'isica Te\'orica, (UAM/CSIC), Universidad 
Aut\'onoma de Madrid, Cantoblanco, E-28049 Madrid, Spain
\and Instituto de Astrof\'isica de Andaluc\'ia (CSIC), Glorieta 
de la Astronom\'ia, E-18008 Granada, Spain}

   \date{}

 
  \abstract{
We report the detection, with a high level of confidence, 
of coherent outflows around voids found in the seventh data release of 
the Sloan Digital Sky Survey (SDSS DR7). In particular, we developed a 
robust $<|{\rm cos}~\theta|>$ statistical test to quantify the strength of 
redshift-space distortions (RSD) associated with extended coherent 
velocity fields.
We consistently find that the vector that joins void centers with galaxies 
that lie in shells around them is more likely to be perpendicular to the 
line-of-sight than parallel to it. This effect is clear evidence 
for the existence of outflows in the vicinity of voids. 
We show that the RSD exist for a wide range of void radius 
and shell thickness, but they are more evident in the largest voids 
in our sample. For instance, we find that the $<|{\rm cos}~\theta|>$ 
for galaxies located in shells within $2 \mpch$ from the edge of 
voids larger than $15 \mpch$ deviates $3.81\sigma$ from uniformity.
The measurements presented in this work provide useful information 
to constrain cosmological parameters, in particular $\Omega_{m}$ 
and $\sigma_{8}$.}

   \keywords{Cosmology: observations - large-scale structure of Universe - 
Galaxies: statistics - Methods: statistical}

   \maketitle
%

\section{Introduction}

According to the cosmological principle, the clustering of galaxies is 
statistically isotropic. In actual galaxy redshift surveys, anisotropies 
arise because distances to galaxies are obtained from redshifts (i.e. Hubble 
flow plus peculiar velocities), making the line-of-sight a preferred 
direction.
These so-called redshift-space distortions (RSD) pose a problem to 
several large-scale structure (LSS) studies, especially those that rely 
on the 3-d positions of galaxies (see \citealt{Hamilton98} for a comprehensive review of 
the effect). Yet, RSD provide valuable information 
about cosmological parameters, in particular the matter density 
parameter $\Omega_{m}$ and the normalization of the amplitude of density 
fluctuations $\sigma_{8}$ (e.g. \citealt{Kaiser87,Cole94,Tinker06, Percival08, 
Ross11}).
The RSD studies have improved considerably since early galaxy surveys thanks to 
measurements made in recent large-scale surveys both from the analysis 
of the galaxy correlation functions (\citealt{Peacock01, Hawkins03} 
using the 2dFGRS and \citealt{Zehavi05,Okumura08,Cabre09} in the SDSS) 
and power spectrum (\citealt{Percival04} in the 2dFGRS). 
These measurements have also been extended to higher redshift \citep{Guzzo08, Blake10}.   
However, it is still debated which the best RSD estimator is for 
extracting cosmological information (see \citealt{Percival08}
for a review of this topic).

A particularly interesting case for observing and quantifying RSD associated 
with extended coherent velocity fields may be provided by {\it large} 
cosmic voids in the distribution of galaxies. 
These regions in the LSS have, on average, a very low enclosed matter density 
($\delta \equiv \rho(<R_{void})-\rho_{mean})/\rho_{mean} \simeq -0.9$,  
e.g. \citealt{Stefan03,Patiri06a}), which generates a well-defined spherically 
averaged radial outflow in the matter (and galaxy) velocity field  
(e.g. \citealt{Padilla05,Ceccarelli06,Patiri06b}). 
As a consequence, the average density contours around large voids that 
are nearly spherical in real space will appear as ellipsoids elongated 
along the line-of-sight in redshift space. That is, galaxies mapped in 
actual redshift surveys that are located in spherical shells around voids 
are more likely to have a position vector perpendicular to the line-of-sight 
than parallel to it. Quantifying this effect provides a powerful 
statistic to assess the strength of RSD. 
Moreover, a key advantage of studying RSD around large voids is that 
the strength of the effect can be computed efficiently for a given power 
spectrum using an existing theoretical framework \citep{Patiri06a, Beta09}, 
which is essential to establish strong constraints to cosmological parameters.

In this paper, we use a sample of galaxies located in the vicinity of large 
voids drawn from the SDSS DR7 to study redshift-space distortions. 
To quantify RSD, we employed a modified form of a particularly 
simple, yet robust method: 
the $<|{\rm cos}~\theta|>$ statistics, i.e. the averaged absolute value of the 
cosine of the angle, $\theta$, between the position vector of galaxies within 
shells around voids and the line-of-sight to the center of their corresponding 
void. This provides a clear, powerful statistic to assess the strength of 
RSD. 

The paper is organized as follows. In section 2 we present the data and 
sample selection. In section 3 we describe the 
statistical test and show our results, and finally in section 4 we discuss 
the results and future work.

\section{Data and sample selection}
\label{sec:samples}

We used the NYU Value-Added Galaxy Catalog 
(NYU-VAGC\footnote{http://sdss.physics.nyu.edu/vagc}, \citealt{Blanton05}) 
associated to the Sloan Digital Sky Survey Data Release Seven 
(SDSS DR7, \citealt{abazajian09}), which is 
the final data release for the SDSS-II. The full catalog 
includes redshifts of $\sim700,000$ main sample galaxies covering 
$\sim8000$ deg$^2$ on the sky. From this catalog, we extracted 
a volume-limited sample defined by a $r$-band absolute magnitude 
threshold $\mrlogh = -20.3$ and a maximum redshift of $z=0.13$, which 
resulted in a sample with a total of 162,076 galaxies. Note that all 
galaxies are $k$-corrected to redshift $z=0.1$ 
(approx. the median redshift of the survey, see \citealt{Blanton03}).

We assumed a flat $\Lambda$CDM cosmology with parameters consistent 
with the latest WMAP-7 results (\citealt{Jarosik}). We converted galaxy 
redshifts to proper comoving distances using the standard equation 
(\citealt{Hogg}) 

\begin{equation}
D_{c}=\frac{c}{H_{0}}\int_{0}^{z}\frac{dz'}{\sqrt{\Omega_{m}(1+z')^{3}+\Omega_{\Lambda}}}, \label{eq:eq1}
\end{equation}
with $\Omega_{m}=0.267$ and $\Omega_{\Lambda}=0.733$. We write the Hubble constant 
$H_{0}$ in terms of the dimensionless Hubble parameter $h$, i.e. $H_{0}=100h$.
We checked that galaxies are uniformly distributed within the sample 
applying a standard $V_{max}$ test.

In this work, we define voids as maximal spheres empty of galaxies brighter 
than the limiting magnitude of our galaxy sample defined above. To search 
for voids in the galaxy sample, we used the {\em HB} void finder. We refer 
the reader to \citet{Patiri06b} for details about the void-finding 
methodology and algorithm. 
In Table \ref{table:t1} we present the statistics of voids found in our 
volume-limited galaxy sample.

Here, we are interested in the galaxies located in the boundaries 
of voids. In particular, we searched for all galaxies located in shells of 
a given width around the edges of the voids. Then, we computed the angle, 
$\theta$, between the position vector of the galaxy (with respect to 
the center of the void) and the position of the center of the void as follows

\begin{equation}
\theta=\cos^{-1}\left(\frac{{\bf s}\cdot{\bf r}}{|{\bf s}||{\bf r}|}\right), \label{eq:eq3}
\end{equation}
where ${\bf s}$ is the position vector of the galaxy with respect to the center 
of the void, and ${\bf r}$ is the position vector of the center of the void with 
respect to the observer.

\begin{table}
\begin{center}
\caption{Complete statistics of large voids found in our volume-limited galaxy sample 
extracted from the SDSS DR7. $N_{V}$ denotes the number of voids larger than the given radius.}
\label{table:t1} 
\begin{tabular}{cr}  

\hline 
Radius & $N_{V}$  \\ 
($\mpch$) &    \\
\hline 

10  & 1162 \\
11  &  829  \\
12  &  464  \\
13  &  284  \\
14  &  141  \\
15  &   76  \\
16  &   34  \\ 
17  &   20  \\
\hline 

\end{tabular}
\end{center}
\end{table}

\section{Statistical test and results}

\begin{table*}
\begin{center}
\caption{Summary of our results.}  
\label{table:t2} 
\begin{tabular}{lcccccccc}  

\hline  \hline
&Radius & $\ll|{\rm cos}~\theta|\gg$ & $rms$ & $N_{\sigma}$ & $p$ & $N_{g}$ & $<|{\rm cos}~\theta|>_{r}$ & $rms_{r}$ \\ 
&($\mpch$) &  &  &  &  & &   &\\
\hline  \hline

Shell width=2.0$\mpch$ & & &  & & & & & \\
\hline               
&10 &  0.4886 &3.758E-03 & 3.02 & $1.047^{+0.016}_{-0.016}$ & 16794 & 0.4991 & 4.7810E-03 \\
&12 &  0.4934 &6.059E-03 & 1.09 & $1.027^{+0.025}_{-0.025}$ &  6926 & 0.4973 & 6.8347E-03 \\
&15 &  0.4509 &1.288E-02 & 3.81 & $1.218^{+0.062}_{-0.065}$ &  1120 & 0.4940 & 1.6674E-02 \\
\hline              
Shell width=4.0$\mpch$ & & & & & & & &  \\
\hline               
&10 &  0.4919 &3.001E-03 & 2.70 & $1.033^{+0.012}_{-0.012}$ &  41837 & 0.4990 & 3.4634E-03 \\
&12 &  0.4935 &4.856E-03 & 1.34 & $1.026^{+0.020}_{-0.020}$ &  17297 & 0.4964 & 5.7295E-03 \\
&15 &  0.4613 &1.123E-02 & 3.44 & $1.168^{+0.052}_{-0.054}$ &   2791 & 0.4903 & 1.3100E-02 \\
\hline              
Shell width=6.0$\mpch$ & & & & & & & &  \\
\hline 
&10 &  0.4949 & 2.593E-03 & 1.96 & $1.021^{+0.011}_{-0.011}$ & 76887 & 0.4999 & 2.8861E-03 \\  
&12 &  0.4933 & 4.182E-03 & 1.60 & $1.027^{+0.017}_{-0.017}$ & 32062 & 0.4986 & 4.5821E-03 \\
&15 &  0.4651 & 1.005E-02 & 3.47 & $1.150^{+0.045}_{-0.047}$ &  5243 & 0.4899 & 9.5722E-03 \\  
\hline 
Shell width=8.0$\mpch$  & & & & & & & &  \\
\hline               
&10 &  0.4966 & 2.331E-03 & 1.46 & $1.014^{+0.009}_{-0.009}$ & 121031 & 0.5000 & 2.5342E-03 \\
&12 &  0.4927 & 3.749E-03 & 1.96 & $1.030^{+0.015}_{-0.016}$ &  51286 & 0.4977 & 4.2074E-03 \\
&15 &  0.4711 & 8.666E-03 & 3.33 & $1.123^{+0.038}_{-0.040}$ &   8520 & 0.4900 & 9.9732E-03 \\

\hline 

\end{tabular}
\tablefoot{$\ll|{\rm cos}~\theta|\gg$ is the average over all voids of the average 
of $|{\rm cos}~\theta|$ of the galaxies in the shells of the void sample larger 
than the given radius. $N_{\sigma}$ is the number of {\it sigmas} for the 
statistics. $N_{g}$ is the total number of galaxies in the given shell. $p$ 
is the {\em anisotropy} parameter. $<|{\rm cos}~\theta|>_{r}$ shows the 
value of $|{\rm cos}~\theta|$ for galaxies in the shells of randomly placed 
spheres and $rms_{r}$ is its root mean square calculated over 100 realizations. 
See text for details.}  
\end{center}
\end{table*}

In principle, variations of standard methods, such as the void-galaxy 
correlation function (see e.g. \citealt{Padilla05}) can be used to detect 
and quantify RSD around voids. However, in the largest 
currently available galaxy survey such as the SDSS, the number of voids and 
galaxies in their vicinity are not sufficient to obtain sharp density contours 
and consequently reliable statistics necessary to extract inferences 
for the relevant cosmological parameters.  
As mentioned above, we developed a modified version of the widely 
used $<|{\rm cos}~\theta|>$ statistical test.

The standard way to construct a $<|{\rm cos}~\theta|>$ test would be to 
simply calculate the average of $|{\rm cos}~\theta|$ over all galaxies in the 
sample located in the relevant shells of a given width around the edges of 
the voids. 
However, because of the correlations between the galaxies within a shell of a 
given void, the corresponding cosine values are not independent random variables. 
This hinders assessing the statistical significance of the test. 
For instance, those correlations may cause the $rms$ to increase considerably 
from the standard $1/\sqrt{12N_{g}}$ $rms$ (where $N_{g}$ is the total number 
of galaxies in the void shells) obtained if galaxies were uncorrelated.

We constructed a modified version of the standard average of the 
$|{\rm cos}~ \theta|$ test in the following way 

\begin{equation}
\ll|{\rm cos}~ \theta|\gg=\frac{1}{N_{V}}\sum_{i=1}^{N_{V}} <|{\rm cos}~ \theta|>_{i},  \label{eq:eq3a}
\end{equation}
where $<|{\rm cos}~ \theta|>_{i}$ is the average of $|{\rm cos}~ \theta|$ over 
all galaxies in the relevant shell of the $i$-th void and $N_{V}$ is the 
total number of voids in our sample. 

In this procedure, galaxies located in the shell of a given void are 
distributed independently for different voids, so they can be treated 
effectively as an independent random variable.
To estimate the $rms$ of our statistic we used the independence of all terms 
of the sum in the r.h.s. of equation \ref{eq:eq3a}, so the variance of our statistic is 
given by

\begin{eqnarray}
rms(\ll|{\rm cos}~ \theta|\gg)^{2} = \frac{1}{N_{V}} \times &\\ 
  \times \left[\frac{\left(\sum_{i=1}^{N_{V}} <|{\rm cos}~ \theta|>^{2}_{i}\right)}{N_{V}}-\ll|{\rm cos}~ \theta|\gg^{2}\right]. & \\ \nonumber
\end{eqnarray}

Once we obtained the $rms$ we computed the statistical significance (i.e. 
giving the confidence level at which the null-hypothesis may be rejected) 
written as the number of {\it sigmas}, $N_{\sigma}$, in the following way

\begin{equation}
\\
N_{\sigma} \equiv \frac{(\ll|{\rm cos}~ \theta|\gg - 0.5)}{rms(\ll|{\rm cos}~ \theta|\gg)}. \\ \label{eq:eq2}
\end{equation}

It is worth noting that this is a conservative approach. In principle, it 
should be possible to construct a more efficient statistical test. This task, 
however, is not simple to carry out since the sampling errors of that 
estimate are difficult to treat. We will address these issues in detail in a 
future work.   

In Table 2 we present the results for the $\ll|{\rm cos}~ \theta|\gg$ 
statistic and its respective significance. 
We show the results obtained for several shell sizes and void radii. In 
absence of redshift distortions, i.e. when the redshift space coincides with 
real space, the average value of the absolute value of ${\rm cos}~\theta$ must 
be $0.5$. 

We double-checked this by placing random spheres over the galaxy sample and 
calculating the angles for the galaxies found in the shells of the random 
spheres following equation \ref{eq:eq3a}. In particular, we made 100 
different realizations of $N_{V}$ random spheres, where $N_{V}$ is the number 
of actual voids larger than a given radius. We then calculated the standard 
average and $rms$ of the 100 realizations. We show these results also in 
Table 2 (columns 6 and 7). Note the similarity between the 
$rms$ values shown in columns 3 and 7. This reflects the fact that the corresponding 
settings, i.e. the number and correlations of galaxies in the 
shells of voids and the random spheres, are rather similar. 
Also, because the values of $<|{\rm cos}~\theta|>$ for the random 
spheres are 0.5 (within the {\em rms}) possible biases caused by 
border effects or by not using the ``correct" cosmological parameters in 
equation \ref{eq:eq1} can be excluded.

For the wide range of void radii and shell sizes considered, the statistic 
generally shows values of $\ll|{\rm cos}~ \theta|\gg$ consistently lower than $0.5$. 
This departure of the distribution of $\ll|{\rm cos}~ \theta|\gg$ from 
uniformity is a clear signal of RSD generated by the 
underdensity associated with the voids. The effect is a unique function 
of the value of the enclosed fractional density contrast ($\delta$) within 
the average radius of the galaxies within the shell, and does not depend 
directly on the value of that radius. 
Moreover, the largest deviations appear for the largest voids in our sample. 
This is not surprising because the average underdensity within 
voids increases with the radius of the void (see e.g. \citealt{Patiri06a}), 
and as mentioned above, if the enclosed density within voids decreases, 
the radial outflows will be stronger.

On the other hand, the effect is more pronounced for the smaller shell sizes. This 
trend can be interpreted by checking the profiles of the enclosed density 
inside and around the voids. Several works 
(e.g. \citealt{Stefan03,Patiri06b,Ceccarelli06}) showed that the void density profiles 
are monotonically increasing. Consequently, for a given sample of 
void radii the effect must be monotonically decreasing (safe for statistical 
fluctuations) as the width of the shell increases.
We obtain, in general, considerable values for $N_{\sigma}$ both for large and 
smaller voids, but the largest {\it sigma} is found for galaxies in shells of 
$2.0 \mpch$ of voids larger than $15.0 \mpch$. 

To show our results in a more intuitive way, we use an {\em anisotropy} 
parameter, {\em p}, defined by
\begin{equation}
p=\frac{1}{\ll|{\rm cos}~ \theta|\gg}-1. \label{eq:eq7}
\end{equation}
Within this parameterization, $p$ values lower than 1 mean that the vectors ${\bf s}$ and ${\bf r}$ 
tend to be parallel, while for $p$ values higher than 1 the vectors tend to be perpendicular.
$p=1$ means compatible with isotropy. We also show the values of $p$ for the cases 
considered above in Table 2, showing more patently the trends seen with 
the $\ll|{\rm cos}~ \theta|\gg$ test. 

\begin{figure}
\centering
\includegraphics[width=\columnwidth]{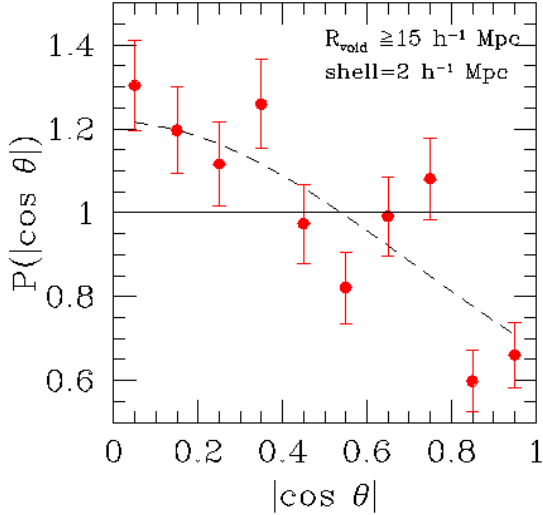}
\caption{Probability distribution of $|{\rm cos}~\theta|$ for voids larger than $15\mpch$ and galaxies 
located in shells of $2\mpch$ (filled circles). The full flat line indicates the distribution for isotropy. The dashed line is a theoretical prediction, see text for details.}
\label{fig:fig1}
\end{figure}

It is also interesting to analyze the probability distribution of the $|{\rm cos}~ \theta|$ values, defined as ${\bf P}(|{\rm cos}~ \theta|)$ (hereafter ${\bf P}$, for brevity). If there are no distortions, ${\bf P}$ must be flat, equal to $1$. In figure \ref{fig:fig1} we show as an example the ${\bf P}$ distribution for voids larger than $15\mpch$ and 
galaxies within shells of $2\mpch$ (filled circles). We can clearly see that the distribution has a declining trend, incompatible with isotropy. 
In recent theoretical works (\citealt{Cuesta08,BT09}), it has been shown that a general 
probability distribution for quadrupolar distortions is adequately described by the following equation:  
\begin{equation}
{\bf P}(|{\rm cos}~\theta|)~d|{\rm cos}~\theta|=\frac{p~d|{\rm cos}~\theta|}{(1+(p^{2}-1)|{\rm cos}~\theta|^{2})^{3/2}},
\end{equation}
where $p$ is the anisotropy parameter defined in equation (\ref{eq:eq7}). We also show  
in figure \ref{fig:fig1} the predicted probability distribution (dashed line) for the case studied above, i.e. with $p=1.218$, which agrees well with the observational results.

\section{Discussion and conclusions}

Our results show clear evidence for the existence of outflows in the vicinity of large 
cosmic voids found in the SDSS DR7. 
Because $\ll|{\rm cos}~ \theta|\gg$ is smaller than $0.5$ for the wide 
range of void radii and shell sizes computed in our sample of galaxies and voids, 
the vector that joins the galaxies located in the shells around voids 
with the centers of those voids is more likely to be perpendicular to the 
line-of-sight than in a random distribution. 
This effect is understood by means of redshift-space distortions: in redshift space, 
our spherical shells intersect higher density isocontours in the direction perpendicular 
to the line-of-sight (recall that a sphere in real space 
is elongated along the line-of-sight due to redshift distortions). Furthermore, 
the effect increases for void with larger radius (for a given shell width), 
while for a fixed void radius it decreases for a broader shell width. 

We may interpret this in terms of the relationship between the deformation of 
the density contour, $T$, and the underlying density contrast, $\delta$, which 
can be written as follows

\begin{equation}
T \propto f(\delta)H_{0}\frac{{\rm d~ln}D}{{\rm d~ln}a}, \label{eq:eqX}
\end{equation}
where $D$ and $a$ are the growth and scale factors, respectively, $H_{0}$ is the 
Hubble constant and $f(\delta)$ is a certain function of the enclosed 
fractional density fluctuation $\delta$ within the radius $R$ \citep{ST02,Beta06}. 
This function monotonically decreases with the absolute value of $\delta$. Therefore, 
for a fixed shell width, the effect is more significant for larger 
voids (since they have higher values of $\delta$). For a fixed void radius, the 
effect is roughly proportional to $f(\delta)$, with $\delta$ equal to the enclosed 
fractional density fluctuation within the mean radius of the shell. Thus, as the 
shell width increases, the effect decreases since the average value 
of $|\delta|$ will be lower. 

In a forthcoming paper, we will explore in detail the model predictions for the 
magnitude of the effect of redshift distortions around voids as a function of 
cosmological parameters. For this we will use the analytical framework 
presented in \citet{Patiri06a} and extended in \citet{Beta09}. In particular, we 
will study the dependence of the $p$ parameter with $\Omega_{m}$ and $\sigma_{8}$. 
The dependence with $\Omega_{m}$ enters through ${\rm d~ln}D/{\rm d~ln}a$ (roughly proportional to $\Omega^{0.6}_{m}$) and also via $\Gamma\equiv\Omega_{m}h$, which characterizes the shape of the power spectrum. The dependence on $\sigma_{8}$ is 
the result of the higher amplitude of the power spectrum, which makes large density 
fluctuations more likely, which in turn increases the value of $|\delta|$.

\begin{acknowledgements}

We thank Ignacio Trujillo and Jordi Miralda-Escud\'e 
for encouraging discussions. We would also like to thank 
the referee, whose comments helped to greatly improve the 
original manuscript. S.G.P. is supported by 
CONICET grant PIP-11420-0148 and by the 
{\it Agencia de Promoci\'on Cient\'ifica y tecnol\'ogica} 
(FONCYT) under contract PICT-2010-2397.
F.P. acknowledges support by MICINN grants AYA2010-21231-C02-01, 
the MultiDark Consolider-Ingenio Program CSD2009-00064 
and the Campus of International Excellence UAM+CSIC.

Funding for the SDSS and SDSS-II has been provided by the 
Alfred P. Sloan Foundation, the Participating Institutions, 
the National Science Foundation, the U.S. Department of Energy, 
the National Aeronautics and Space Administration, the Japanese 
Monbukagakusho, the Max Planck Society, and the Higher Education 
Funding Council for England. The SDSS Web Site is http://www.sdss.org/.

The SDSS is managed by the Astrophysical Research Consortium 
for the Participating Institutions. The Participating Institutions 
are the American Museum of Natural History, Astrophysical Institute 
Potsdam, University of Basel, University of Cambridge, Case Western 
Reserve University, University of Chicago, Drexel University, Fermilab, 
the Institute for Advanced Study, the Japan Participation Group, Johns 
Hopkins University, the Joint Institute for Nuclear Astrophysics, the 
Kavli Institute for Particle Astrophysics and Cosmology, the Korean 
Scientist Group, the Chinese Academy of Sciences (LAMOST), Los Alamos 
National Laboratory, the Max-Planck-Institute for Astronomy (MPIA), 
the Max-Planck-Institute for Astrophysics (MPA), New Mexico State 
University, Ohio State University, University of Pittsburgh, University 
of Portsmouth, Princeton University, the United States Naval Observatory, 
and the University of Washington.

\end{acknowledgements}

\end{document}